\documentclass[showpacs,twocolumn,pre]{revtex4}
\usepackage{amsmath}
\usepackage{latexsym}
\usepackage{epsfig}
\usepackage{makeidx}
\usepackage{amsfonts}

\begin{document}

\title{Single integro-differential wave equation for L\'{e}vy walk }
\author{Sergei Fedotov }
\affiliation{School of Mathematics, The University of Manchester, Manchester M13 9PL, UK}

\begin{abstract}
The integro-differential wave equation for the probability density function
for a classical one-dimensional L\'{e}vy walk with continuous sample paths
has been derived. This equation involves a classical wave operator together
with memory integrals describing the spatio-temporal coupling of the L\'{e}%
vy walk. It is valid for any running time PDF of a walker and it does not
involve any long-time large-scale approximations. It generalizes the
well-known telegraph equation obtained from the persistent random walk.
Several non-Markovian cases are considered when the particle's velocity
alternates at the gamma and power-law distributed random times.
\end{abstract}

\maketitle

L\'{e}vy walk is a fundamental notion in physics and biology \cite{KlSo}
with numerous applications including the transport of light governed by L%
\'{e}vy statistics \cite{La}, anomalous superdiffusion of cold atoms in
optical lattices \cite{Ba}, T-cell motility in the brain \cite{Le2},
endosomal active transport within living cells \cite{Gr}. In the last few
years, the interest in the L\'{e}vy walk models increases rapidly in the
context of L\'{e}vy flight foraging hypothesis according to which a L\'{e}vy
transport is optimal to search for randomly located objects \cite{Le4}. L%
\'{e}vy movement pattern has been observed in microorganisms, insects,
molluscs, birds, etc. \cite{Men}. Recent publication \textit{L\'{e}vy walk}
in\textit{\ Review of Modern Physics} provides a detailed discussion of
existing applications and the most current status of L\'{e}vy walk theory
\cite{Za}.

The derivation of the governing equations for a L\'{e}vy walk from an
underlying stochastic movement is a long-standing problem. Many
contributions on this subject have been given since the pioneering works
three decades ago (see, for example, \cite{Kl,Blu}). The standard model for
a L\'{e}vy walk is based on the continuos time random walk (CTRW) with
coupled probability density function for the jump length and the waiting
time between two successive jumps. Two integral equations for the
probability density function (PDF) $p(x,t)$ for a walker position $x$ at
time $t$ and arrival rate $j\left( x,t\right) $ can be formulated and solved
by the Fourier-Laplace transform technique \cite{KlSo,Za}. An equivalent
single integral equation for $p(x,t)$ has been also formulated \cite{Zabur}.
To describe a L\'{e}vy walk spatiotemporal coupling, two dynamical equations
involving fractional material derivatives have been suggested in \cite{SokR}%
. Another approach to describe the superdiffusive behavior is based on the
analysis of joint PDF $p(x,v,t)$ of the particle's position $x$ and velocity
$v$. Various fractional generalizations of Kramers-Fokker-Planck equation
for $p(x,v,t)$ \ have been derived \cite{Kra1,Kra2,Kra3}. However, to the
author's knowledge, the single integro-differential equation for the L\'{e}%
vy walk PDF $p(x,t)$ is still not available in the literature. In this paper
we obtain the following integro-differential wave equation
\begin{equation*}
\frac{\partial ^{2}p}{\partial t^{2}}-v^{2}\frac{\partial ^{2}p}{\partial
x^{2}}+\int_{0}^{t}\int_{V}K\left( \tau \right) \varphi (u)\left( \frac{%
\partial }{\partial t}-u\frac{\partial }{\partial x}\right) \times
\end{equation*}%
\begin{equation}
p\left( x-u\tau ,t-\tau \right) dud\tau =0,  \label{main}
\end{equation}%
where $v$ is a constant speed of walker, $\varphi (u)$ is the velocity jump
density:%
\begin{equation}
\varphi (u)=\frac{1}{2}\delta \left( u-v\right) +\frac{1}{2}\delta \left(
u+v\right)  \label{jump}
\end{equation}%
in the velocity space $V$ and $K(\tau )$ is the standard memory kernel from
the CTRW theory \cite{KlSo}. It is determined by its Laplace transform $\hat{%
K}(s)=\hat{\psi}(s)/\hat{\Psi}(s),$ where $\hat{\psi}(s)$ and $\hat{\Psi}(s)$
are the Laplace transforms of the running time density $\psi (\tau )$ and\
the survival function $\Psi (\tau )$. From Eq.\ (\ref{main}) with $%
p(x,0)=p_{0}\left( x\right) $ and $p_{t}(x,0)=0,$ one can obtain the
well-known expression for the Fourier-Laplace transform of the PDF $p(x,t)$
\begin{equation}
\hat{p}(k,s)=\frac{\left[ \hat{\Psi}(s+ikv)+\hat{\Psi}(s-ikv)\right] \hat{p}%
_{0}(k)}{2-\hat{\psi}(s+ikv)-\hat{\psi}(s-ikv)},  \label{LF}
\end{equation}%
where $\hat{p}_{0}(k)=\int_{\mathbb{R}}p_{0}\left( x\right) e^{ikx}dx$ \cite%
{Za}.

\textit{Derivation.} We consider the L\'{e}vy walk as the random particle's
motion with continuos sample paths (no jumps) along one-dimensional space.
The particle starts to move with constant speed $v$ at time $t=0$ and after
a random time (running time) it either continues the movement in the same
direction with curtain probability or changes the direction and moves with
the same constant speed. The random running time is defined by the switching
rate $\gamma (\tau )$ or the running time PDF $\psi (\tau )=\gamma (\tau
)\exp \left[ -\int_{0}^{\tau }\gamma (s)ds\right] .$ To derive the governing
equation for the L\'{e}vy walk, we start with Markovian model involving
structural densities with the extra running time variable $\tau $ \cite%
{MFH,Cox,Vlad,Fer,FedotAbby}. We define the structural PDF's of walker, $%
n_{+}(x,t,\tau ),$ at point $x$ and time $t$ that moves in the right
direction, $\left( +\right) ,$ with constant speed $v$ during time $\tau $
since the last switching. The probability density function $n_{-}(x,t,\tau )$
corresponds to the walker that moves in the negative direction, $\left(
-\right) .$ The balance equations for both structural PDF's $n_{+}(x,t,\tau
) $ and $n_{-}(x,t,\tau )$ can be written as
\begin{equation}
\frac{\partial n_{\pm }}{\partial t}\pm v\frac{\partial n_{\pm }}{\partial x}%
+\frac{\partial n_{\pm }}{\partial \tau }=-\gamma (\tau )n_{\pm },
\label{str}
\end{equation}%
where the switching rate $\gamma (\tau )$ depends on the running time $\tau $%
. If the walker moves in the positive direction it can switch with rate $%
\gamma (\tau )$ either to the opposite direction with the probability $%
\alpha _{-}$ or keep the same direction with the probability $\alpha _{+}$
such that $\alpha _{+}+\alpha _{-}=1.$\ For the walker moving in the
negative direction corresponding characteristics are $\beta _{+}$ and $\beta
_{-}.$ The well-known \textit{velocity model} and \textit{two-state model}
are just particular cases of this general two-state model. For example, the
choice $\alpha _{+}=\beta _{-}=0$ and $\alpha _{-}=\beta _{+}=1$ corresponds
to the two-state model \cite{Mas}. The probabilities $\alpha _{\pm }=\beta
_{\pm }=1/2$ correspond to the velocity model \cite{Zu}. We assume that at
the initial time $t=0$ all walkers start to move with zero running time $%
\tau $
\begin{equation}
n_{\pm }(x,0,\tau )=\frac{1}{2}p_{0}(x)\delta (\tau ).  \label{initial}
\end{equation}%
Here we consider the symmetrical initial conditions when walker start to
move on the right with probability $1/2$ and on the left with the same
probability. The boundary conditions at zero running time can be formulated
as
\begin{eqnarray}
n_{\pm }(x,t,0) &=&\alpha _{\pm }\int_{0}^{t}\gamma (\tau )n_{+}(x,t,\tau
)d\tau +  \notag \\
&&\beta _{\pm }\int_{0}^{t}\gamma (\tau )n_{-}(x,t,\tau )d\tau .
\label{zero2}
\end{eqnarray}%
Our aim is to derive the master equations for the probability density
functions $p_{+}(x,t)$ and $p_{-}(x,t)$ defined as%
\begin{equation}
p_{\pm }(x,t)=\int_{0}^{t}n_{\pm }(x,t,\tau )d\tau .  \label{den}
\end{equation}%
By differentiating (\ref{den}) with respect to time $t$ and using the
balance equations (\ref{str}) we obtain
\begin{eqnarray*}
\frac{\partial p_{\pm }}{\partial t} &=&n_{\pm }(x,t,t)\mp v\int_{0}^{t}%
\frac{\partial n_{\pm }}{\partial x}d\tau \\
&&-\int_{0}^{t}\frac{\partial n_{\pm }}{\partial \tau }d\tau
-\int_{0}^{t}\gamma (\tau )n_{\pm }(x,t,\tau )d\tau .
\end{eqnarray*}%
This equation can be rewritten as the master equation%
\begin{equation}
\frac{\partial p_{\pm }}{\partial t}\pm v\frac{\partial p_{\pm }}{\partial x}%
=j_{\pm }(x,t)-i_{\pm }(x,t),  \label{mas}
\end{equation}%
where the rates of switching $i_{\pm }(x,t)$ are defined as
\begin{equation}
i_{\pm }(x,t)=\int_{0}^{t}\gamma (\tau )n_{\pm }(x,t,\tau )d\tau  \label{i0}
\end{equation}%
and the arrival rates $j_{\pm }(x,t)$ are
\begin{equation}
j_{\pm }(x,t)=n_{\pm }(x,t,0).  \label{ar}
\end{equation}%
By using the definitions of switching and arrival rates (\ref{i0}) and (\ref%
{ar}), Eq. (\ref{zero2}) can be written in the compact form%
\begin{equation}
j_{\pm }(x,t)=\alpha _{\pm }i_{+}(x,t)+\beta _{\pm }i_{-}(x,t).  \label{ra}
\end{equation}%
In what follows we consider only the simple case of a symmetric L\'{e}vy
walk for which $\alpha _{\pm }=\beta _{\pm }=1/2.$ In general, the
probabilities $\alpha _{\pm }$ and $\beta _{\pm }$ can be useful to
formulate the impact of the external force or chemotactic substance.
Substitution of (\ref{ra}) with $\alpha _{\pm }=\beta _{\pm }=1/2$\ into the
master equation (\ref{mas}) gives
\begin{equation}
\frac{\partial p_{+}}{\partial t}+v\frac{\partial p_{+}}{\partial x}=-\frac{1%
}{2}i_{+}(x,t)+\frac{1}{2}i_{-}(x,t),  \label{mean3}
\end{equation}%
\begin{equation}
\frac{\partial p_{-}}{\partial t}-v\frac{\partial p_{-}}{\partial x}=\frac{1%
}{2}i_{+}(x,t)-\frac{1}{2}i_{-}(x,t),  \label{mean4}
\end{equation}%
where the switching rates $i_{\pm }(x,t)$ can be found as follows. By the
method of characteristics, we find from (\ref{str})
\begin{equation}
n_{\pm }(x,t,\tau )=n_{\pm }(x\mp v\tau ,t-\tau ,0)\Psi (\tau ),\quad \tau
<t,  \label{cha}
\end{equation}%
where $\Psi (\tau )$ is the survival function
\begin{equation}
\Psi (\tau )=e^{-\int_{0}^{\tau }\gamma (s)ds}.  \label{surva}
\end{equation}%
Note that at $\tau =t$ we have a singularity due to the initial condition (%
\ref{initial}). Substitution of (\ref{cha}) into (\ref{i0}) and (\ref{den})
together with the initial condition (\ref{initial}) gives
\begin{equation*}
i_{\pm }(x,t)=\int_{0}^{t}j_{\pm }(x\mp v\tau ,t-\tau )\psi (\tau )d\tau +%
\frac{1}{2}p_{0}(x\mp vt)\psi (t),
\end{equation*}%
\begin{equation*}
p_{\pm }(x,t)=\int_{0}^{t}j_{\pm }(x\mp v\tau ,t-\tau )\Psi \left( \tau
\right) d\tau +\frac{1}{2}p_{0}(x\mp vt)\Psi \left( t\right) .
\end{equation*}%
Applying the Fourier-Laplace transform to these equations, we can find
expressions for $i_{+}(x,t)$ and $i_{-}(x,t)$ in terms of $\rho _{+}(x,t)$
and $\rho _{-}(x,t)$ as \cite{FedotAbby}
\begin{equation}
i_{\pm }(x,t)=\int_{0}^{t}K(\tau )p_{\pm }(x\mp v\tau ,t-\tau )d\tau .
\label{rate_ii}
\end{equation}%
Note that four integral equations $i_{\pm }(x,t)$ and $p_{\pm }(x,t)$ above
can be written in the standard form of two equations for the PDF:%
\begin{equation}
p(x,t)=p_{+}(x,t)+p_{-}(x,t).  \label{total}
\end{equation}%
and the arrival rate $j=j_{+}+j_{-}$ with the jump density $w(z|\tau )=\frac{%
1}{2}\delta \left( z-v\tau \right) +\frac{1}{2}\delta \left( z+v\tau \right)
.$ For the symmetrical case $j=i_{+}+i_{-}$ we obtain the well-known
equations%
\begin{eqnarray*}
p(x,t) &=&\int_{0}^{t}\int_{\mathbb{R}}j(x-z,t-\tau )w(z|\tau )\Psi \left(
\tau \right) dzd\tau + \\
&&\int_{\mathbb{R}}p_{0}(x-z)w(z|t)dz\Psi \left( t\right) ,
\end{eqnarray*}%
\begin{eqnarray*}
j(x,t) &=&\int_{0}^{t}\int_{\mathbb{R}}j(x-z,t-\tau )w(z|\tau )\psi (\tau
)dzd\tau + \\
&&\int_{\mathbb{R}}p_{0}(x-z)w(z|t)dz\psi (t).
\end{eqnarray*}%
\bigskip Note that these two equations are the starting point for the
various studies of the L\'{e}vy walk \cite{Za}.

\bigskip Our main purpose is to reduce the system (\ref{mean3}), (\ref{mean4}%
) with (\ref{rate_ii}) to a single governing equation for the PDF $p(x,t)$
defined by (\ref{total}). First we introduce the flux \cite{Kac}
\begin{equation}
J(x,t)=vp_{+}(x,t)-vp_{-}(x,t).  \label{flux}
\end{equation}%
Then by adding (\ref{mean3}) and (\ref{mean4}) we obtain the standard
conservation equation
\begin{equation}
\frac{\partial p}{\partial t}+\frac{\partial J}{\partial x}=0.  \label{ro}
\end{equation}%
Equation for the flux $J$ can be obtain by multiplication of (\ref{mean3})
and (\ref{mean4}) by $v$ and subtraction
\begin{equation}
\frac{\partial J}{\partial t}+v^{2}\frac{\partial p}{\partial x}=-v\left[
i_{+}(x,t)-i_{-}(x,t)\right] .  \label{J}
\end{equation}%
By differentiating (\ref{ro}) with respect to $t$ and (\ref{J}) with respect
to $x$ and eliminating $\partial ^{2}J/\partial t\partial x,$ we obtain%
\begin{equation}
\frac{\partial ^{2}p}{\partial t^{2}}=v^{2}\frac{\partial ^{2}p}{\partial
x^{2}}+v\frac{\partial }{\partial x}\left[ i_{+}(x,t)-i_{-}(x,t)\right] .
\label{kkk}
\end{equation}%
Now we need to express the last term in (\ref{kkk}) in terms of $p(x,t)$
alone. From (\ref{total}) and (\ref{flux}) we find the expressions for $%
p_{+}(x,t)$ and $p_{-}(x,t)$ in terms of\ the PDF $p(x,t)$ and the flux $%
J(x,t)$
\begin{equation}
p_{\pm }(x,t)=\frac{p(x,t)}{2}\pm \frac{J(x,t)}{2v}.  \label{+}
\end{equation}%
Substitution of (\ref{+}) into (\ref{rate_ii}) gives
\begin{equation*}
v\frac{\partial }{\partial x}\left[ i_{+}(x,t)-i_{-}(x,t)\right] =
\end{equation*}%
\begin{eqnarray}
&&\frac{1}{2}\int_{0}^{t}K\left( \tau \right) [\frac{\partial J}{\partial x}%
(x-v\tau ,t-\tau )+\frac{\partial J}{\partial x}(x+v\tau ,t-\tau )  \notag \\
&&+v\frac{\partial p}{\partial x}(x-v\tau ,t-\tau )-v\frac{\partial p}{%
\partial x}(x+v\tau ,t-\tau )]d\tau .  \label{iiii}
\end{eqnarray}%
By using (\ref{iiii})\ together with (\ref{ro}) and (\ref{kkk}), we obtain a
single integro-differential equation for the PDF $p(x,t)$
\begin{equation*}
\frac{\partial ^{2}p}{\partial t^{2}}=v^{2}\frac{\partial ^{2}p}{\partial
x^{2}}-
\end{equation*}%
\begin{eqnarray}
&&\frac{1}{2}\int_{0}^{t}K\left( \tau \right) \left[ \left( \frac{\partial }{%
\partial t}-v\frac{\partial }{\partial x}\right) p\left( x-v\tau ,t-\tau
\right) \right] d\tau -  \notag \\
&&\frac{1}{2}\int_{0}^{t}K\left( \tau \right) \left[ \left( \frac{\partial }{%
\partial t}+v\frac{\partial }{\partial x}\right) p\left( x+v\tau ,t-\tau
\right) \right] d\tau .  \label{cla}
\end{eqnarray}%
This integro-differential wave equation for a \textit{L\'{e}vy walk }is
valid for any running time PDF $\psi \left( \tau \right) $. It generalizes
the well-known telegraph equation obtained from the persistent random walk
with the constant rate of switching $\gamma .$ By using the velocity jump
density (\ref{jump}), we rewrite Eq. (\ref{cla}) in the compact form (\ref%
{main}). It is instructive to compare the uncoupled CTRW with L\'{e}vy walk.
When jumps and waiting times of the CTRW are independent random variables,
one can convert the single integral equation for $p(x,t)$ into a master
equation \cite{KlSo,MFH}. For the L\'{e}vy walk involving spatio-temporal
coupling the integro-differential wave equation (\ref{cla}) plays the same
role as the master equation for the uncoupled CTRW.

Let us now consider several examples of the running time PDF $\psi \left(
\tau \right) .$

\textit{Exponential running time density. }In the Markovian case with the
exponential running time PDF
\begin{equation}
\psi \left( \tau \right) =\ \frac{1}{T}\exp \left( -\frac{\tau }{T}\right)
\label{expo}
\end{equation}%
for which $\hat{\psi}(s)=\left( 1+Ts\right) ^{-1}$ and $K\left( \tau \right)
=T^{-1}\delta \left( \tau \right) $, we obtain from (\ref{cla}) the
classical Cattaneo or telegraph equation \cite{MFH}
\begin{equation}
\frac{\partial ^{2}p}{\partial t^{2}}+\frac{1}{T}\frac{\partial p}{\partial t%
}-v^{2}\frac{\partial ^{2}p}{\partial x^{2}}=0.  \label{Cat}
\end{equation}%
This hyperbolic equation ensures that the density profile propagates with
finite speed $v.$

\textit{Gamma PDF }$g(\tau ,2,\lambda ).$\textit{\ }For the biological
applications it is important to consider a running time PDF that takes the
maximum value not at zero time as the exponential density (\ref{expo}) \cite%
{Men}. For example, it was found that the running time density for a single
bacterium might deviate significantly from exponential approximation \cite%
{Kor}. One example of such PDF is the \textit{gamma density} \textit{\ }%
\begin{equation}
\psi \left( \tau \right) =g(\tau ,2,\lambda )=\ \lambda ^{2}\tau \exp \left(
-\lambda \tau \right)  \label{Ga}
\end{equation}%
with \ $\hat{\psi}(s)=\lambda ^{2}$ $\left( s+\lambda \right) ^{-2}$ and $%
\hat{K}(s)=\lambda ^{2}$ $\left( 2\lambda +s\right) ^{-1}.$ The memory
kernel in Eq. (\ref{cla}) takes the form
\begin{equation}
K\left( \tau \right) =\lambda ^{2}\exp \left( -2\lambda \tau \right) .
\end{equation}%
The advantage of this exponential memory kernel is that one can localize
integro-differential equation (\ref{cla}) by direct differentiation of (\ref%
{cla}) with respect to time twice:
\begin{equation*}
\frac{\partial ^{4}p}{\partial t^{4}}+4\lambda \frac{\partial ^{3}p}{%
\partial t^{3}}+5\lambda ^{2}\frac{\partial ^{2}p}{\partial t^{2}}+2\lambda
^{3}\frac{\partial p}{\partial t}=
\end{equation*}%
\begin{equation}
v^{2}\frac{\partial ^{2}}{\partial x^{2}}\left[ 2\frac{\partial ^{2}p}{%
\partial t^{2}}+4\lambda \frac{\partial p}{\partial t}+3\lambda ^{2}p-v^{2}%
\frac{\partial ^{2}p}{\partial x^{2}}\right] .  \label{four}
\end{equation}%
Note that non-Markovian particle's movement with velocities alternating at
Erlang-distributed and gamma-distributed random times was considered in \cite%
{Ital0,Ital}. Next we consider the anomalous case involving walker's
velocities alternating at power-law distributed random times \cite%
{SokR,Fer,Mas}.

\textit{Anomalous enhanced transport.} We consider two anomalous cases: (1)
strong ballistic case for which the mean squared displacement: $<x^{2}>\sim
t^{2}$ and (2) subballistic superdiffusion with $<x^{2}>\sim t^{3-\mu },$
where $1<\mu <2$ \cite{Za}. In\ the ballistic case, we use the survival
function $\Psi \left( \tau \right) =E\left[ -\left( \tau /\tau _{0}\right)
^{\mu }\right] $ with $0<\mu <1$ for which the mean running time is
divergent \cite{Sca,Steve}. In this case
\begin{equation}
\hat{\psi}\left( s\right) =\frac{1}{1+\left( \tau _{0}s\right) ^{\mu }}%
,\qquad 0<\mu <1
\end{equation}%
and the Laplace transform of the memory kernel $K\left( \tau \right) $ is
\begin{equation*}
\hat{K}(s)=\frac{s\hat{\psi}(s)}{1-\hat{\psi}(s)}=\frac{s^{1-\mu }}{\tau
_{0}^{\mu }}.
\end{equation*}%
For this kernel the main equation (\ref{main}) can be rewritten in the
different forms by using material fractional derivatives \cite%
{SokR,Mer0,Bae,Mer1,Sib,Mag}. We write it in the form
\begin{equation}
\frac{\partial ^{2}p}{\partial t^{2}}-v^{2}\frac{\partial ^{2}p}{\partial
x^{2}}+\frac{1}{\tau _{0}^{\mu }}\int_{V}\varphi (u)L_{u}^{1-\mu }pdu=0,
\label{lee}
\end{equation}%
where the operator $L_{u}^{1-\mu }$ is defined by its Fourier-Laplace
transform
\begin{equation}
\mathcal{FL}\left\{ L_{u}^{1-\mu }p\right\} =\left( s-iku\right) ^{1-\mu }%
\left[ \left( s+iku\right) p(k,s)-p_{0}(k)\right] .  \label{O}
\end{equation}%
In the subballistic superdiffusive case, one can obtain for small $s$
\begin{equation}
\hat{\psi}\left( s\right) \simeq 1-Ts+ATs^{\mu },\qquad 1<\mu <2  \label{A}
\end{equation}%
for which the first moment $T=\int_{0}^{\infty }\Psi (\tau )d\tau $ is
finite and the second moment is divergent. Then
\begin{equation*}
\hat{K}(s)\simeq \frac{1}{T}\left( 1+As^{\mu -1}\right)
\end{equation*}%
as $s\rightarrow 0.$ From (\ref{main}) we obtain the following equation
\begin{equation}
T\frac{\partial ^{2}p}{\partial t^{2}}+\frac{\partial p}{\partial t}-D\frac{%
\partial ^{2}p}{\partial x^{2}}+A\int_{V}\varphi (u)L_{u}^{1-\mu }pdu=0
\label{le}
\end{equation}%
with the diffusion coefficient $D=Tv^{2}.$ It is clear from (\ref{lee}) and (%
\ref{le}) that ballistic and subballistic cases are fundamentally different.
In the strong ballistic case ($0<\mu <1$), the integral term is in the
balance with classical wave equation, while for the subballistic
superdiffusive case ($1<\mu <2$), the memory term is in the balance with the
Cattaneo (telegraph) equation. One can perform various asymptotic analysis
of (\ref{lee}) and (\ref{le}), obtain the pseudo-differential equations for
the walker's PDF position \cite{Bae,Mer1,Sib,Mag} and determine the shape of
PDF profiles \cite{Bar}.

As an illustration let us find the long-time asymptotic solution to Eq. (\ref%
{lee}). In the limit $\tau _{0}\rightarrow 0$, the evolution of the PDF $%
p\left( x,t\right) $ is determined by the integral term, while the first two
wave equation terms in (\ref{lee}) are irrelevant. The PDF $p\left(
x,t\right) $ obeys
\begin{equation}
\int_{V}\varphi (u)L_{1-\mu }^{u}pdu=0.  \label{La}
\end{equation}%
By using (\ref{O}), we obtain
\begin{equation*}
\left( s-ikv\right) ^{1-\mu }\left[ \left( s+ikv\right) \hat{p}(k,s)-\hat{p}%
_{0}(k)\right] +
\end{equation*}%
\begin{equation}
\left( s+ikv\right) ^{1-\mu }\left[ \left( s-ikv\right) \hat{p}(k,s)-\hat{p}%
_{0}(k)\right] =0.  \label{LL}
\end{equation}%
In particular, for $\mu =1/2$ and $\hat{p}_{0}(k)=1$, we factorize\ Eq. (\ref%
{LL}) and find
\begin{equation*}
\left( s^{2}+k^{2}v^{2}\right) ^{\frac{1}{2}}\hat{p}(k,s)-1=0.
\end{equation*}%
By using the inverse Fourier-Laplace transform we obtain the well-known
self-similar profile \cite{KlSo}
\begin{equation*}
p\left( x,t\right) =\pi ^{-1}\left( v^{2}t^{2}-x^{2}\right) ^{-\frac{1}{2}}.
\end{equation*}%
For the arbitrary $\mu $ from the interval $0<\mu <1,$ one can find from (%
\ref{LL}) that the solution to (\ref{La}) is the Lamperti distribution \cite%
{Za}. It would be interesting to use a new wave equation for the
non-normalizable density problem for superdiffusive anomalous transport \cite%
{Re}. Note that our equation can be generalized for the case when the
particle moves with the random velocity: $\dot{x}_{\pm }(t)=\pm v+\dot{B}%
\left( t\right) ,$ where $\dot{B}\left( t\right) $ is the Gaussian white
noise.

In summary, we derived the integro-differential wave equation for the
probability density function for a position of a L\'{e}vy walker with
continuous sample paths. This equation involves a classical wave operator
together with memory integrals describing a spatio-temporal coupling of L%
\'{e}vy walk. It is valid for any running time PDF and it does not involve
any long-time large-scale approximations. It generalizes the well-known
telegraph equation obtained from the persistent random walk. For the L\'{e}%
vy walk the integro-differential wave equation (\ref{main}) plays the same
role as the master equation for the uncoupled CTRW \cite{KlSo,MFH}. Our
technique may lead to the significant advances in the extension of the
linear L\'{e}vy walk models. Our approach may be particularly helpful when
we deal with the \textit{L\'{e}vy particles }interactions \cite{Int} when it
would be difficult to take into account nonlinear terms within the standard
approach involving integral equations (see the similar problems for
subdiffusion \cite{F2,F1}). The description of the \textit{L\'{e}vy walk} in
terms of the Markovian balance equations (\ref{str}) proves to be very
useful for the analysis of \textit{L\'{e}vy walk }with a random death
process \cite{FedotAbby}. Our theory can be useful to formulate the impact
of the external force or the chemotactic substance on the random movement of
particles with finite velocities. This theory can be also useful for the
implementation of the non-linear reactions and development of the theory of
wave propagation in reaction-transport systems involving enhanced diffusion
and memory effects \cite{MFH,Memory}.

\textit{Acknowledgement}. This work was funded by EPSRC Grant No.
EP/J019526/1 \textit{Anomalous reaction-transport equations}. The author
gratefully acknowledges the hospitality of the Isaac Newton Institute,
Cambridge and thanks Nickolay Korabel and Vicenc Mendez for very useful
discussions during the programme \textit{Coupling Geometric PDEs with
Physics for Cell Morphology, Motility and Pattern Formation. } The author
thanks Eli Barkai for useful suggestions.

\end{document}